\documentclass[aps,graphicx]{revtex4}
\usepackage{graphicx}
\usepackage{amsmath}

\begin{document}


\title{Universal distributed quantum computing on superconducting qutrits with dark photons\footnote{Published in Ann. Phys. (Berlin)  \textbf{530}, 1700402 (2018).}}

\author{Ming Hua$^{1,2,3}$, Ming-Jie Tao$^{1}$, Ahmed Alsaedi$^{2}$, Tasawar  Hayat$^{2,4}$,
and Fu-Guo Deng$^{1,2,}$\footnote{Corresponding
author:fgdeng@bnu.edu.cn} }

\address{$^{1}$ Department of Physics, Applied Optics Beijing Area Major Laboratory,
Beijing Normal University, Beijing 100875, China\\
$^{2}$ NAAM-Research Group, Department of Mathematics, Faculty of
Science, King Abdulaziz University, Jeddah 21589, Saudi Arabia\\
$^{3}$ Department of Applied Physics, School of Science, Tianjin
Polytechnic University, Tianjin 300387, China\\
$^{4}$ Department of Mathematics, Quaid-I-Azam University, Islamabad
44000, Pakistan}

\begin{abstract}
We present a one-step scheme to construct the controlled-phase gate
deterministically on remote transmon qutrits coupled to different
resonators connected by a superconducting transmission line for an
universal distributed quantum computing. Different from previous
works on remote superconducting qubits, the present gate is
implemented with coherent evolutions of the entire system in the
all-resonance regime assisted by the dark photons to robust against
the transmission line loss, which allows the possibility of the
complex designation of a long-length transmission line to link lots
of circuit QEDs. The length of the transmission line can reach the
scale of several meters, which makes our scheme is suitable for the
large-scale distributed quantum computing. This gate is a fast
quantum entangling operation with a high fidelity of about $99\%$.
Compare with previous works in other quantum systems for a
distributed quantum computing, under the all-resonance regime, the
present proposal does not require classical pulses and ancillary
qubits, which relaxes the difficulty of its implementation largely.
\end{abstract}


\keywords{Distributed quantum computing; superconducting qutrits;
dark photons; quantum information; Quantum electrodynamics}

\maketitle


\section{Introduction}

Quantum computation (QC) \cite{Nielsen,longguilu}, as an
interdisciplinary research of computer science and quantum
mechanics, has attracted much attention in recent years. It can
implement the famous Shor's algorithm \cite{shor} for the
factorization of an n-bit integer exponentially faster than the
classical algorithms and the Grover's algorithm \cite{Grover} or the
optimal Long's algorithm \cite{LongGrover} for unsorted database
search. Various quantum systems have been used to implement  QC,
such as  photons
\cite{Knill,photon,hyperCNOT2,hyperCNOT3,hyperCNOT4,hyperCNOT5,photon2},
nuclear magnetic resonance \cite{NMR,Long1,NHQCLong}, diamond
nitrogen-vacancy center
\cite{Togan,NVgate,weiNVgate,songxk1,songxk2,cai}, and cavity
quantum electrodynamics (QED) \cite{book}. Among the quantum
systems, circuit QED \cite{Blais,Wallraff}, composed of a
superconducting qubit (SQ) coupled to a superconducting resonator
(SR), provides a good platform for the implementation of  QC because
of its good ability of the large-scale integration and the accurate
manipulation on the SQ \cite{Barends,zhengyuan}.

Circuit QED has been studied a lot for achieving the basic tasks of
QC on SQs or SRs, such as the construction of the single-qubit and
the universal quantum gates
\cite{DiCarlo,Haack,Strauch,Hua1,Hua2,McKay,HPaik}, entangled state
generation
\cite{Steffen,Cao,Leghtas,Strauch1,Strauch2,Aron,Felicetti,Narla,Koshino},
and the measurement and the non-demolition detection on SQs or SRs
\cite{AWallraff,Johnson,Feng,hantianyi}. The types for integrating the SQs and
the SRs mainly contain some SQs coupled to a SR bus \cite{Majer} or
some SRs coupled to a SR bus \cite{Hua3} or a SQ
\cite{Wu,Yang,YangCP}. At present, it is hard to integrate lots of
SQs or SRs in a quatum-bus-based processor to achieve the complex
universal QC. Further scaling up the number of SQs or SRs requires
linking the distant circuit QED systems to form a quantum network
\cite{Pellizzari,Loo,YY,SJ,MP,Roch,Mandt,sheng,qiang,zhoulan1,zhoulan2,zhoulan3,yanga} introduced by the
\emph{distributed quantum computing} \cite{Cirac}, in which a
quantum computer can be seen as a quantum network of distant local
processors with only a few qubits and are connected by quantum
transmission lines (TL). As the key problem in the realization of
the distributed quantum computing, quantum entanglement and
universal quantum gate on remote qubits have been discussed in some
other systems \cite{Clark,Browne,Duan1,Mancini,dengfuguo,yanhui}. For example, Cirac
\emph{et al.} \cite{Cirac1} proposed a scheme to achieve the ideal
quantum transmission between atoms trapped at spatially separated
nodes in 1997. In 2004, Xiao \emph{et al} \cite{Xiao} realized the
controlled phase (c-phase) gate between two rare-earth ions embedded
in the respective microsphere cavities assisted by a single-photon
pulse in sequence. In 2011, L\"u \emph{et al} \cite{Lu} proposed two
schemes to complete the entanglement generation and quantum-state
transfer between two spatially separated semiconductor quantum dot
molecules.

To achieve the universal quantum gate on distant qubits coupled to
different cavities connected by the TLs\cite{Yin1,Clader,Yin2,Sai},
realistic flying-photon qubit or adiabatic processes and the local
operations are required.  On one hand, there are some works which
studied the quantum network by using the dark photon in the TL in
other quantum systems. In 2007, Yin \emph{et al} \cite{Yin1}
presented some schemes to achieve the state transfer and quantum
entangling  gates deterministically between the remote multiple
two-level atoms trapped in different cavities connected by an
optical fiber, in which the c-phase gate should be completed by
using the ``dipole blockade" effect among atoms in a cavity and it
needs not to populate the realistic photons in the fiber. In 2014,
Clader \cite{Clader} presented an adiabatic scheme to transfer a
microwave quantum state from one cavity to another, assisted by an
optical fiber which is robust against both mechanical and fiber
loss. On the other hand, one should transfer the microwave photon to
the optical photons to link the remote SQs. In 2015, Yin \emph{et
al.} \cite{Yin2} proposed a scheme to achieve the quantum networking
of SQs based on the optomechanical interface.

To implement the distributed quantum computing on remote SQs coupled
to different SRs connected by a superconducting TL, one should
overcome the decay of the TL as the more the complicated designation
and a longer length for the TL is,  the bigger the decay of the
photon in it becomes. In this paper, we propose a scheme for the
construction of the c-phase gate on two remote transmon qutrits
coupled to different SRs connected by a superconducting TL for the
distributed quantum computing on SQs. Our scheme works in the
all-resonance regime by letting the frequencies of the qutrits and
the resonators equal to each other. The scheme can be achieved with
just one step assisted by the dark photons in the TL, without
requiring classical pulses and ancillary qubits, which relaxes the
difficulty of its implementation in experiment largely. Far
different from the c-phase gate on two remote superconducting
resonators constructed in Ref.\cite{Hua2} which is completed with
three resonance steps between resonators and a qubit and can be
extended to achieve the gate on two remote superconducting qubits by
coupling them to the two remote resonators, respectively, we use a
superconducting TL instead of the superconducting qubit as a quantum
bus. Here, using the dark photons in TL to reduce the requirement of
the quality factor of the TL allows the complex designation of a TL
to link lots of remote circuit QEDs and the length of the TL (the
distance between two remote superconducting qubits) can reach the
scale of several meters. The fidelity of the present c-phase gate is
beyond $99\%$ by using the numerical simulation with the feasible
parameters.

\section{Basic theories}

\begin{figure}[tpb]
\begin{center}
\includegraphics[width=10.0cm,angle=0]{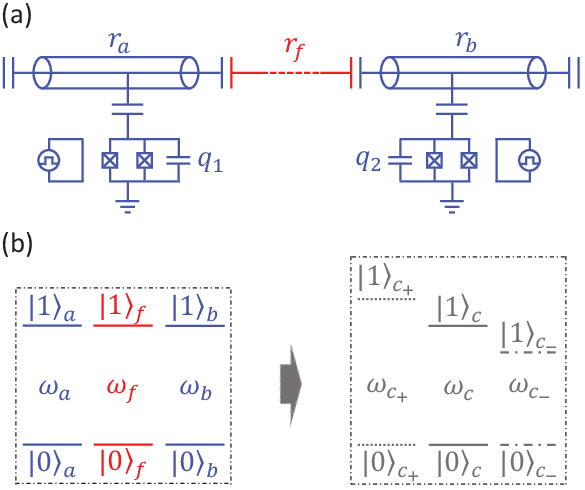}
\end{center}
\caption{(a) Setup for a distributed quantum computing  composed of
two remote qubits $q_1$ and $q_2$ coupled to different resonators
$r_a$ and $r_b$ connected by a transmission line $r_f$. (b)
Illustrations of the energy splitting of the subsystem composed of
$r_a$, $r_b$, and $r_f$.} \label{fig1}
\end{figure}

Let us consider a distributed quantum computing composed of two
remote superconducting qubits $q_1$ and $q_2$ coupled to two
single-mode high-quality superconducting resonators $r_a$ and $r_b$,
respectively, which are connected by a superconducting TL $r_f$,
shown in Fig.~\ref{fig1}. The Hamiltonian of this device is (in the
interaction picture with $\hbar=1$)
\begin{eqnarray}             
H  &=&  H_1^a +H_2^b + H_f^{a(b)}\nonumber\\
&=&  g_{1}^{a} (a\!^{+} \sigma_{1}^{-} e^{-i\delta_{1}^{a}t}
 + a\sigma_{1}^{+}e^{i\delta_{1}^{a}t})  +\, g_{2}^{b}(b\!^{+} \sigma_{2}^{-}e^{-i\delta_{2}^{b}t}
 + b\sigma_{2}^{+}e^{i\delta_{2}^{b}t}) \nonumber\\
&&  + \sum_{j=1}^{\infty} g_{f,j}^{I} \left[f_{j}\,(a^{+} +
(-1)^{j}e^{i\phi}b^{+}) + H.c. \right], \label{initial}
\end{eqnarray}
where  $H_1^a$, $H_2^b$, and $H_f^{a(b)}$ are the interaction
Hamiltonians of the subsystems composed of $q_1$ and $r_a$, $q_2$
and $r_b$, and $r_f$ and $r_a$ ($r_b$), respectively. $H_f^{a(b)}$
applies to the high-finesse resonators and resonant operations over
the time scale much longer than the TL's round-trip time \cite{Enk}.
$\delta_{J}^{I}=\omega_{I}-\omega_{J}$ ($I=a$,$b$ and
$J=1$,$2$,$f$). $\omega_a$, $\omega_b$, and $\omega_f$ are the
transition frequencies of resonators $r_a$, $r_b$, and the TL $r_f$,
respectively. $\omega_1$ and $\omega_2$ are the transition
frequencies of the qubits $q_1$ and $q_2$, respectively. $a^{+}$,
$b^{+}$, and $f^{+}$ are the creation operators of the resonators
$r_a$, $r_b$, and the TL $r_f$, respectively. $\sigma _{1}^{+}$ and
$\sigma _{2}^{+}$ are the creation operators of the transitions
$|g\rangle_{1}\leftrightarrow|e\rangle_1$ and
$|g\rangle_{2}\leftrightarrow|e\rangle_2$ of the qubits $q_{1}$ and
$q_{2}$, respectively. $|g\rangle_{1(2)}$ and $|e\rangle_{1(2)}$ are
the ground and the first excited states of the qubit $q_{1(2)}$,
respectively. $g_{1}^{a}$ and $g_{2}^{b}$ are the coupling strength
between $q_1$ and $r_a$ and that between $q_2$ and $r_b$,
respectively. $g_{f,j}^{I}$ is the coupling strength between
$r_{a(b)}$ and the mode $j$ of the TL $r_f$. $\phi$ is the phase
induced by the propagating field through the TL $r_f$ of length $l$
with the relation $\phi=2\pi\omega l/c$ in which $c$ is the speed of
light.

In the short TL limit $(2L\kappa_{f}^{a(b)})/(2\pi c)\leq 1$, only
one resonant mode $f$ of the TL $r_f$ interacts with the resonators'
modes ($L$ is the length of $r_f$  and  $\kappa_{f}^{a(b)}$ is the
decay rate of the resonator $r_{a(b)}$ into a continuum of TL modes)
\cite{Yin1}. The Hamiltonian $H$ can be reduced to
\begin{eqnarray}             
H_{int} &=&  g_{1}^{a}(a^{+}\sigma_{1}^{-}e^{-i\delta_{1}^{a}t}
 + a\sigma_{1}^{+}\!e^{i\delta_{1}^{a}t}) +\, g_{2}^{b}(b^{+} \sigma_{2}^{-}e^{-i\delta_{2}^{b}t}
 + a\sigma_{2}^{+}\!e^{i\delta_{2}^{b}t})   +\, g_{f}^{a}(f^{+}a + fa^{+}) + g_{f}^{b}(f^{+}b + fb^{+}).
\label{cphamiltonian}
\end{eqnarray}
In the Schr\"odinger picture, this Hamiltonian can be rewritten as
\begin{eqnarray}             
H'  &=&   \omega_{a} a^{+}a + \omega_{b} b^{+}b + \omega_{f} f^{+}f
+ \omega_{1}\sigma _{1}^{+}\sigma _{1}^{-}
+ \omega_{2}\sigma _{2}^{+}\sigma _{2}^{-} \nonumber\\
&&   +\, g_{1}^{a}(a^{+}\, \sigma _{1}^{-} + a \sigma _{1}^{+}) +
g_{2}^{b}(b^{+}\, \sigma _{2}^{-} + b \sigma _{2}^{+})
  +\, g_{f}^{a}(f^{+}\,a + f\,a^{+}) + g_{f}^{b}(f^{+}\,b +
f\,b^{+}). \label{cphamiltonian01}
\end{eqnarray}
To construct the c-phase gate on the
remote transmon qutrits below, we consider the all-resonance
condition with
$\omega_a=\omega_b=\omega_f=\omega_{1}=\omega_{2}=\omega$ by letting
the frequencies of the qubits and the resonators and the TL  equal
to each other and $g_{f}^{a}=g_{f}^{b}=g$ by letting the coupling
strength between $r_a$ and $r_f$ equal to the one of $r_b$ and
$r_f$. If one takes the canonical transformations
$C_{\pm}=\frac{1}{2}(a+b \pm \sqrt{2}f)$ and
$C=\frac{\sqrt{2}}{2}(a-b)$ \cite{Serafini,WLYang}, the  Hamiltonian
$H'$  can be represented as
\begin{eqnarray}             
H''  &=&   \omega \sigma  _{1}^{+} \sigma_{1}^{-}  + \omega \sigma
_{2}^{+} \sigma_{2}^{-} + \omega C^{+}C +  \left(\omega +
\sqrt{2}g\right)C_{+}C_{+}^{+} + \left(\omega  -
\sqrt{2}g\right)C_{-}C_{-}^{+}   +
\frac{1}{2}\Big[g_{1}^{a}\left(C_{+}  +  C_{-} + \sqrt{2}c\right)\sigma_{1}^{+} \nonumber\\
&&   + g_{1}^{a}\left( C_{+}^{+} +  C_{-}^{+} +
\sqrt{2}C^{+}\right)\sigma_{1}^{-}   + g_{2}^{b}\left( C_{+}  +
C_{-} - \sqrt{2}C\right)\sigma_{2}^{+}  + g_{2}^{b}\left(C_{+}^{+} +
C_{-}^{+}  - \sqrt{2}C^{+}\right)\sigma_{2}^{-}\Big].
\label{cphamiltonian02}
\end{eqnarray}
Here the modes $C$ and $C_{\pm}$ are three bosonic modes and they
are not coupled to each other. From  Eq. (\ref{cphamiltonian02}),
the energy level of the subsystem composed of $r_a$, $r_b$, and
$r_f$ are split  into three different parts with frequencies
$\omega_{c_{+}}$, $\omega_{c_{-}}$, and $\omega_{c}$ signed by the
modes $C_{+}$, $C_{-}$, and $C$, respectively, as shown in Fig.
\ref{fig1} (b). Because of the contributions of the fields of $r_a$
and $r_b$, the three modes $C$ and $C_{\pm}$  interact with the two
qubits $q_1$ and $q_2$. When $g \gg \{g_{1}^{a},g_{2}^{b}\}$, the
excitations of modes $C_{\pm}$ are highly suppressed as $\omega \pm
\sqrt{2}g$ detune with the resonance modes ($C$, $q_1$, and $q_2$
with frequency of $\omega$) largely, which means the modes $C_{\pm}$
are the dark ones to the frequency $\omega_f$ of $r_f$, and the
Hamiltonian $H''$ can be reduced to
\begin{eqnarray}             
H'''  = \omega \sigma _{1}^{+}\sigma _{1}^{-} + \omega
\sigma_{2}^{+}\sigma _{2}^{-} + \omega C^{+}C   +
\frac{1}{\sqrt{2}}\left[g_{1}^{a}\left(C \sigma_{1}^{+} + C^{+}
\sigma_{1}^{-}\right) - g_{2}^{b}\left(C \sigma_{2}^{+} +
C^{+}\sigma_{2}^{-}\right)\right]. \label{cphamiltonian03}
\end{eqnarray}
It can be written as
\begin{eqnarray}             
H_{e\!f\!f} = \frac{1}{\sqrt{2}}\left[g_{1}^{a}(C \sigma_{1}^{+} +
C^{+}\sigma_{1}^{-}) - g_{2}^{b}(C \sigma_{2}^{+} +
C^{+}\sigma_{2}^{-})\right] \label{effect}
\end{eqnarray}
in the interaction picture. Here, only the mode
$C=\frac{\sqrt{2}}{2}(a-b)$ is left, which means that the TL can not
be populated in the all-resonance regime in our system. The
interaction between two remote two-energy-level qubits expressed by
$H_{e\!f\!f}$ will be used to construct the
all-resonance c-phase gate on the two remote qutrits
below.

\begin{figure}[tpb]
\begin{center}
\includegraphics[width=11.0cm,angle=0]{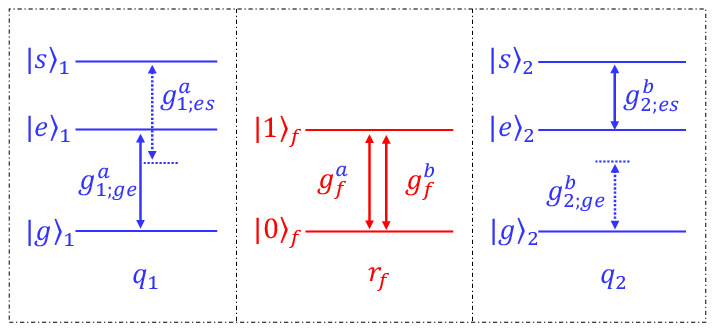}
\end{center}
\caption{Illustrations of interactions between $q_1$ and $r_a$,
$r_f$ and $r_a$ ($r_b$), and $q_2$ and $r_b$, respectively, for the
construction of the c-phase gate on two remote transmon qutrits
$q_1$ and $q_2$.} \label{fig4}
\end{figure}

\section{C-phase gate on the two remote qutrits $q_1$ and $q_2$}

To construct the c-phase gate on the two remote transmon qutrits
$q_1$ and $q_2$ in the device shown in Fig. \ref{fig1}, one should
consider the evolutions of states $|g\rangle_{1}|g\rangle_{2}$,
$|g\rangle_{1}|e\rangle_{2}$, $|e\rangle_{1}|g\rangle_{2}$, and
$|e\rangle_{1}|e\rangle_{2}$ simultaneously, and generate a minus
phase on one of them only. According to Eq. (\ref{effect}) and
considering the two lowest transitions of the qubits only discussed
in the section II, the phase of the state
$|g\rangle_{1}|g\rangle_{2}$ doesn't evolve with time. The phase of
the state $|e\rangle_{1}|e\rangle_{2}$ is the product of the phases
of states $|g\rangle_{1}|e\rangle_{2}$ and
$|e\rangle_{1}|g\rangle_{2}$ which are both evolve with time. There
are three states evolve with time. Based on these relationships
among the phases of the four states, one cannot construct the
one-step all-resonance c-phase gate.

Here, we consider the second excited energy level $|s\rangle_2$ of
$q_2$ and take
$\omega_{1;ge}/(2\pi)=\omega_{2;es}/(2\pi)=\omega_{a}/(2\pi)=\omega_{b}/(2\pi)=\omega_{f}/(2\pi)$.
The illustrations of the interactions between $q_1$ and $r_a$, $r_f$
and $r_a$ ($r_b$), and $q_2$ and $r_b$  for constructing the c-phase
gate on $q_1$ and $q_2$ are shown in Fig. \ref{fig4}. The
Hamiltonian of the whole system can be expressed as
\begin{eqnarray}             
H_{2q}   &=&  g_{1;ge}^{a}
\left(a^{+}\sigma_{1;ge}^{-}e^{-i\delta_{1;ge}^{a}t}+a\sigma_{1;ge}^{+}e^{i\delta_{1;ge}^{a}t}\right)
 + g_{1;es}^{a} \left(a^{+}\sigma_{1;es}^{-}e^{-i\delta_{1;es}^{a}t}+a\sigma_{1;es}^{+}e^{i\delta_{1;es}^{a}t}\right)   \nonumber\\
&&  +  g_{2,ge}^{b}
\left(b^{+}\sigma_{2;ge}^{-}e^{-i\delta_{2;ge}^{b}t}+b\sigma_{2;ge}^{+}e^{i\delta_{2;ge}^{b}t}\right)
 + g_{2;es}^{b} \left(b^{+}\sigma_{2;es}^{-}e^{-i\delta_{2;es}^{b}t}+b\sigma_{2;es}^{+}e^{i\delta_{2;es}^{b}t}\right)   \nonumber\\
&&  +  g_{f}^{a}\left(f^{+}a + fa^{+}\right)  +
g_{f}^{b}\left(f^{+}b + fb^{+}\right), \label{3energy}
\end{eqnarray}
in which $\sigma _{1(2);ge}^{+}$ and $\sigma _{1(2);es}^{+}$ are the
creation operators of the transitions
$|g\rangle_{1(2)}\leftrightarrow |e\rangle_{1(2)}$ and
$|e\rangle_{1(2)}\leftrightarrow |s\rangle_{1(2)}$ of the qutrit
$q_{1(2)}$, respectively. $g_{1(2);ge}^{a(b)}$ and
$g_{1(2);es}^{a(b)}$
$\left(g_{1(2);es}^{a(b)}=\sqrt{2}g_{1(2);ge}^{a(b)}\right)$ are the
coupling strengths between the two transitions of $q_{1(2)}$ and
$r_{a(b)}$, respectively.
$\delta_{1(2);ge}^{a(b)}=\omega_{1(2);ge}-\omega_{a(b)}$ and
$\delta_{1(2);es}^{a(b)}=\omega_{1(2);es}-\omega_{a(b)}$.
$\omega_{1(2);ge}$ ($\omega_{1(2);ef}$) is the frequency of the
transition $|g\rangle_{1(2)} \leftrightarrow |e\rangle_{1(2)}$
$\left(|e\rangle_{1(2)} \leftrightarrow |s\rangle_{1(2)}\right)$ of
the qutrit $q_{1(2)}$. $|s\rangle_{1(2)}$ is the second excited
state of $q_{1(2)}$. In order to get two states of the qubits
evolve with time only, one should take
$\omega_{1(2);ge}-\omega_{1(2);es} \gg
\{g_{1;ge}^{a},g_{2;ge}^{b}\}$ to ignore the dispersive coupling
between the transition $|e\rangle_{1} \leftrightarrow |s\rangle_{1}$
of the qutrit $q_{1}$ and $r_{a}$ and the one between the transition
$|g\rangle_{2} \leftrightarrow |e\rangle_{2}$ of the qutrit $q_{2}$
and $r_{b}$. The Hamiltonian $H_{2q}$ can be reduced to
\begin{eqnarray}             
H_{2q}'   &=&  g_{1;ge}^{a}(a^{+}\sigma_{1;ge}^{-}  +
a\sigma_{1;ge}^{+}) + g_{2;es}^{b}(b^{+} \sigma_{2;es}^{-} +
b\sigma_{2;es}^{+})  +\, g_{f}^{a}(f^{+}a + fa^{+}) +
g_{f}^{b}(f^{+}b + fb^{+}). \label{2q'}
\end{eqnarray}

To achieve the one-step all-resonance c-phase gate by using the dark
photon in the superconducting TL, we take the same canonical
transformations as the ones in Sec. II and $g_{f}^{a}=g_{f}^{b} \gg
\{g_{1;ge}^{a},g_{2;es}^{b}\}$, the Hamiltonian  $H_{2q}'$  becomes
\begin{eqnarray}             
H_{e\!f\!f}' = \frac{1}{\sqrt{2}}\left[g_{1;ge}^{a} (C
\sigma_{1;ge}^{+} + C^{+} \sigma_{1;ge}^{-}) - g_{2;es}^{b}(C
\sigma_{2;es}^{+} + C^{+}\sigma_{2;es}^{-})\right]. \label{effect'}
\end{eqnarray}
Suppose that $|\psi_1\rangle=|g\rangle_1|g\rangle_2|0\rangle_c$,
$|\psi_2\rangle=|g\rangle_1|e\rangle_2|0\rangle_c$,
$|\psi_3\rangle=|e\rangle_1|g\rangle_2|0\rangle_c$, and
$|\psi_4\rangle=|e\rangle_1|e\rangle_2|0\rangle_c$ ($|0\rangle_c
\equiv |0\rangle_a|0\rangle_b|0\rangle_f$) are the initial states of
the system undergoes the Hamiltonian $H_{e\!f\!f}'$, respectively,
one can get their evolutions as
\begin{eqnarray}             
|\Psi_1(t)\rangle =
e^{iH_{e\!f\!f}'t}|g\rangle_1|g\rangle_2|0\rangle_c =
|g\rangle_1|g\rangle_2|0\rangle_c, \label{m1'}
\end{eqnarray}
\begin{eqnarray}     
|\Psi_2(t)\rangle =
e^{-iH_{e\!f\!f}'t}|g\rangle_1|e\rangle_2|0\rangle_c =
|g\rangle_1|e\rangle_2|0\rangle_c, \label{m2'}
\end{eqnarray}
\begin{eqnarray}             
|\Psi_3(t)\rangle  &=&  e^{-iH_{e\!f\!f}'t}|e\rangle_1|g\rangle_2|0\rangle_c  \nonumber\\
&=&
\cos\left(\frac{g_{1;ge}^{a}}{\sqrt{2}}t\right)|e\rangle_1|g\rangle_2|0\rangle_c
+\sin\left(\frac{g_{1;ge}^{a}}{\sqrt{2}}t\right)|g\rangle_1|g\rangle_2|1\rangle_c,
\label{m3'}
\end{eqnarray}
\begin{eqnarray}             
|\Psi_4(t)\rangle  &=&  e^{iH_{ef\!f}'t}|e\rangle_1|e\rangle_2|0\rangle_c  \nonumber\\
&=&  \frac{1}{G'}\left[(g_{2;es}^{b})^{2}
+(g_{1;ge}^{a})^{2}\cos\left(\sqrt{\frac{G'}{2}}t\right)\right]|e\rangle_1|e\rangle_2|0\rangle_c  \nonumber\\
&& -\frac{g_{1;ge}^{a} g_{2;es}^{b}}{G'}\left[\cos\left(\sqrt{\frac{G'}{2}}t\right)-1\right]|g\rangle_1|s\rangle_2|0\rangle_c \nonumber\\
&& -\frac{i
g_{1;ge}^{a}}{\sqrt{G'}}\sin\left(\sqrt{\frac{G'}{2}}t\right)|g\rangle_1|e\rangle_2|1\rangle_c,
\label{m4'}
\end{eqnarray}
where $G'=(g_{1;ge}^{a})^{2}+(g_{2;es}^{b})^{2}$. From the
evolutions of the four states, one can construct the c-phase gate on
the two remote qutrits $q_1$ and $q_2$. In detail, we suppose the
initial state of the system described by $H_{e\!f\!f}'$ is
\begin{eqnarray}             
|\Psi_0^{cp}\rangle = (\cos{\theta_1}|g\rangle_1
+\sin{\theta_1}|e\rangle_1) \otimes(\cos{\theta_2}|g\rangle_2
+\sin{\theta_2}|e\rangle_2) \otimes |0\rangle_c. \label{cp0}
\end{eqnarray}
According to Eqs. (\ref{m1'}) and (\ref{m2'}), one can keep the
states $|g\rangle_1|g\rangle_2|0\rangle_c$ and
$|g\rangle_1|e\rangle_2|0\rangle_c$ unchanged. By taking the proper
$g_{1;ge}^{a}$ and $g_{2;ge}^{b}$ to satisfy
$\frac{g_{1;ge}^{a}}{\sqrt{2}}t=(2k-1)\pi$ and
$\sqrt{\frac{G'}{2}}t=2m\pi$ ($k,m=1,2,3,\cdots$) simultaneously,
one can achieve the condition that when the state
$|e\rangle_1|g\rangle_2|0\rangle_c$ undergoes an odd number of
periods and generates a minus phase (from Eq. (\ref{m3'})), the
state $|e\rangle_1|e\rangle_2|0\rangle_c$ undergoes an even number
of periods and keeps unchanged (from Eq. (\ref{m4'})) meanwhile.
That is, the state of the system evolves from $|\Psi_0^{cp}\rangle$
to the final state
\begin{eqnarray}             
|\Psi_f^{cp}\rangle = (\alpha_1|g\rangle_1|g\rangle_2 +
\alpha_2|g\rangle_1|e\rangle_2 - \alpha_3|e\rangle_1|g\rangle_2 +
\alpha_4|e\rangle_1|e\rangle_2) \otimes |0\rangle_c, \label{cpf}
\end{eqnarray}
which is just the target state after our c-phase gate operation on
$q_1$ and $q_2$ with the initial state $|\Psi_0^{cp}\rangle$. Here
$\alpha_1=\cos{\theta_1}\cos{\theta_2}$,
$\alpha_2=\cos{\theta_1}\sin{\theta_2}$,
$\alpha_3=\sin{\theta_1}\cos{\theta_2}$, and
$\alpha_4=\sin{\theta_1}\sin{\theta_2}$. In the basis
$\{|g\rangle_1|g\rangle_2,|g\rangle_1|e\rangle_2,|e\rangle_1|g\rangle_2,|e\rangle_1|e\rangle_2
\}$, the matrix of the c-phase gate is
\begin{equation}           
U^{cp}=\left(
\begin{array}{cccc}
1 & 0 & 0 & 0 \\
0 & 1 & 0 & 0 \\
0 & 0 & -1 & 0 \\
0 & 0 & 0 & 1
\end{array}
\right).
\end{equation}

\begin{figure}[tpb]
\begin{center}
\includegraphics[width=10.0cm,angle=0]{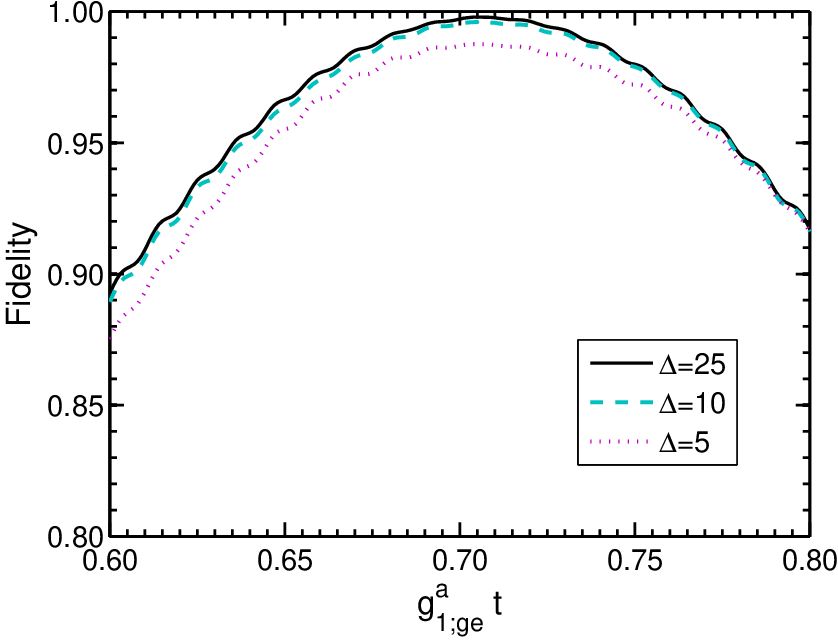}
\caption{The fidelity of our c-phase gate varies with $gt$ and
different $\Delta=g_{f}^{a(b)}/g_{1;ge}^{a}$.}\label{fig5}
\end{center}
\end{figure}

To get the c-phase gate within a short time, we take $k=m=1$, that
is,
$g_{2;es}/(2\pi)=\sqrt{2}g_{2;ge}/(2\pi)=\sqrt{3}g_{1;ge}/(2\pi)$.
Supposing that the initial state of the system is
$|\Psi_{max}\rangle=\frac{1}{2}(|\psi_1\rangle+|\psi_2\rangle+|\psi_3\rangle+|\psi_4\rangle)$,
one can get the state
$|\Psi_{max}^{cp}\rangle=\frac{1}{2}(|\psi_1\rangle+|\psi_2\rangle-|\psi_3\rangle+|\psi_4\rangle)$
after our c-phase gate operation on the two remote qutrits $q_1$ and
$q_2$ with a maximal fidelity of $99.8\%$ ($\Delta=25$), $99.6\%$
($\Delta=10$), and $98.8\%$ ($\Delta=5$) within $gt=0.705$
($\omega_{1;ge}-\omega_{1;es}=\omega_{2;ge}-\omega_{2;es}=90g_{1;ge}^{a}$),
by using the definition
\begin{eqnarray}          
F_{max}^{cp} = |\langle
\Psi_{max}^{cp}|e^{-iH_{2q}t}|\Psi_{max}\rangle|^2, \label{fidelity}
\end{eqnarray}
as shown in Fig. \ref{fig5}. Here $\Delta\equiv
g_{f}^{a(b)}/g_{1;ge}^{a}$. One can tune the frequencies of transmon
qutrits by using individual flux bias lines \cite{Reed}, which let
the frequencies of qutrits resonate or detune with resonators to
turn on or off the operation of our gate.

\section{Possible experimental implementation and fidelity}

In experiment, a high quality factor $Q\sim2\times 10^6$ of a 1D SR
has been demonstrated \cite{Megrant}. By considering the relation
between the decay rate $\kappa$, $Q$, and the frequency of resonator $\omega_r$
with $\kappa=\omega_r/Q$ \cite{Blais}, the best life time of a
photon in a superconducting resonator can reach $\sim50$ $\mu$s. The
coherence time of a transmon qubit \cite{Reed, Koch, Schreier, Chow}
can also reach $50$ $\mu$s by using titanium nitride \cite{Chang}.
The tunable range of the transition frequency of a transmon qubit
can reach $2.5$ GHz, which helps us to tune our transmon qutrits to
resonate (detune) with their corresponding resonator (largely)
effectively. The coupling strength between a transmon qutrit and a
SR can be realized larger than $200$ MHz \cite{Steffen1}. The
anharmonicity between the two transitions of a transmon qutrit can
reach
$\omega_{1;ge}/(2\pi)-\omega_{1;es}/(2\pi)=\omega_{2;ge}/(2\pi)-\omega_{2;es}/(2\pi)=
0.72$ GHz \cite{Hoi}, which lets us  ignore the detune interaction
between each qutrit and their corresponding resonator, compared with
the small coupling strength between them. As for the SRs and the TL
$r_f$, one can couple them by using the SQUID, which can reach a
coupling strength of $g_{f}^{a(b)}/(2\pi) \sim 200$ MHz
theoretically with reasonable experimental parameters
\cite{Peropadre}. Moreover, one can also use the capacitance
coupling between resonators and the superconducting TL. With the
reasonable parameters
$\omega_{a}/(2\pi)=\omega_{b}/(2\pi)=\omega_{f}/(2\pi)= 6$ GHz, the
coupling capacitance $C=13.3$ fF, and the capacitance per unit
length of the transmission line and the resonators $C_r=2$ pF
\cite{Blais,hu}, the capacitance coupling strength can reach
$g_{f}^{a(b)}/(2\pi)=40$ MHz (which will be discussed below for the
construction of the c-phase gate with $\Delta=5$). It is worth
noticing that a coupling strength between a SR and a superconducting
TL has been realized with about $32$ MHz \cite{YY}.

To show the feasibility of our scheme for the construction of the
c-phase gate on two remote qutrits, we numerically simulate the
fidelity of the scheme based on the parameters realized in
experiments or predicted theoretically with reasonable experimental
parameters.

The dynamics of the quantum system undergoes the Hamiltonian
$H_{2q}$ is determined by the master equation
\begin{eqnarray}              
\frac{d\rho }{dt}  &=&  -i[H_{2q},\rho ]+\kappa_a D[a]\rho +\kappa_b
D[b]\rho + \kappa_f D[f]\rho
+ \sum_{l=1,2}\{\gamma_{l;ge}D[\sigma_{l;ge}^{-}]\rho+\gamma_{l;es}D[\sigma_{l;es}^{-}]\rho  \nonumber\\
&& +
\gamma_{l;e}^{\phi}(\sigma_{l;ee}\rho\sigma_{l;ee}-\sigma_{l;ee}\rho/2-\rho
\sigma_{l;ee}/2)  +
\gamma_{l;s}^{\phi}(\sigma_{l;ss}\rho\sigma_{l;ss}-\sigma_{l;ss}\rho/2-\rho
\sigma_{l;ss}/2)\}. \label{masterequation}
\end{eqnarray}
Here, $\kappa_{a,b,f}$ is the decay rate of the resonator
$r_{a,b,f}$. $\gamma_{l;ge}$ ($\gamma_{l;es}$) and
$\gamma_{l;e}^{\phi}$ ($\gamma_{l;s}^{\phi}$) are the energy
relaxation and the dephase rates of the transition $|e\rangle_l
\leftrightarrow |g\rangle_l$ ($|s\rangle_l \leftrightarrow
|e\rangle_l$) of the transmon qutrits $q_l$ ($l=1,2$), respectively.
$\gamma_{l;ge}^{-1}=(\gamma_{l;ge}^{\phi})^{-1}=(\gamma_{l;es}^{\phi})^{-1}=2\gamma_{l;es}^{-1}$
\cite{Strand}, $\sigma_{l;ee}=|e\rangle_l\langle e|$,  and
$\sigma_{l;ss}=|s\rangle_l\langle s|$. $D[L]\rho=(2L\rho
L^{+}-L^{+}L\rho-\rho L^{+}L)/2$. Because of the competition
relation between the coupling strength $g_{1;ge}^{a}$ and the decay
rates and that between the decoherence time of resonators and
qutrits, for different $\gamma_{l;ge}$ and $\kappa_{a,b,f}$, one
should choose different $g_{1;ge}^{a}$ to reach the maximal fidelity
of our scheme for constructing the c-phase gate on remote qutrits
$q_1$ and $q_2$. The coupling strengths
$g_{1;ge}^{a}/(2\pi)$ chosen below
are the optimal ones which correspond to the highest fidelities of
the gate when we fix $g_{f}^{a(b)}/(2\pi)=200$ MHz and
$\gamma_{l;ge}^{-1}=\kappa_{a,b,f}= 50$ $\mu$s by considering the
set of discretized $g_{1;ge}^{a}$ values, varying from 1 to 100MHz
in steps of 1MHz.

\begin{figure}[tpb]
\begin{center}
\includegraphics[width=12.0cm,angle=0]{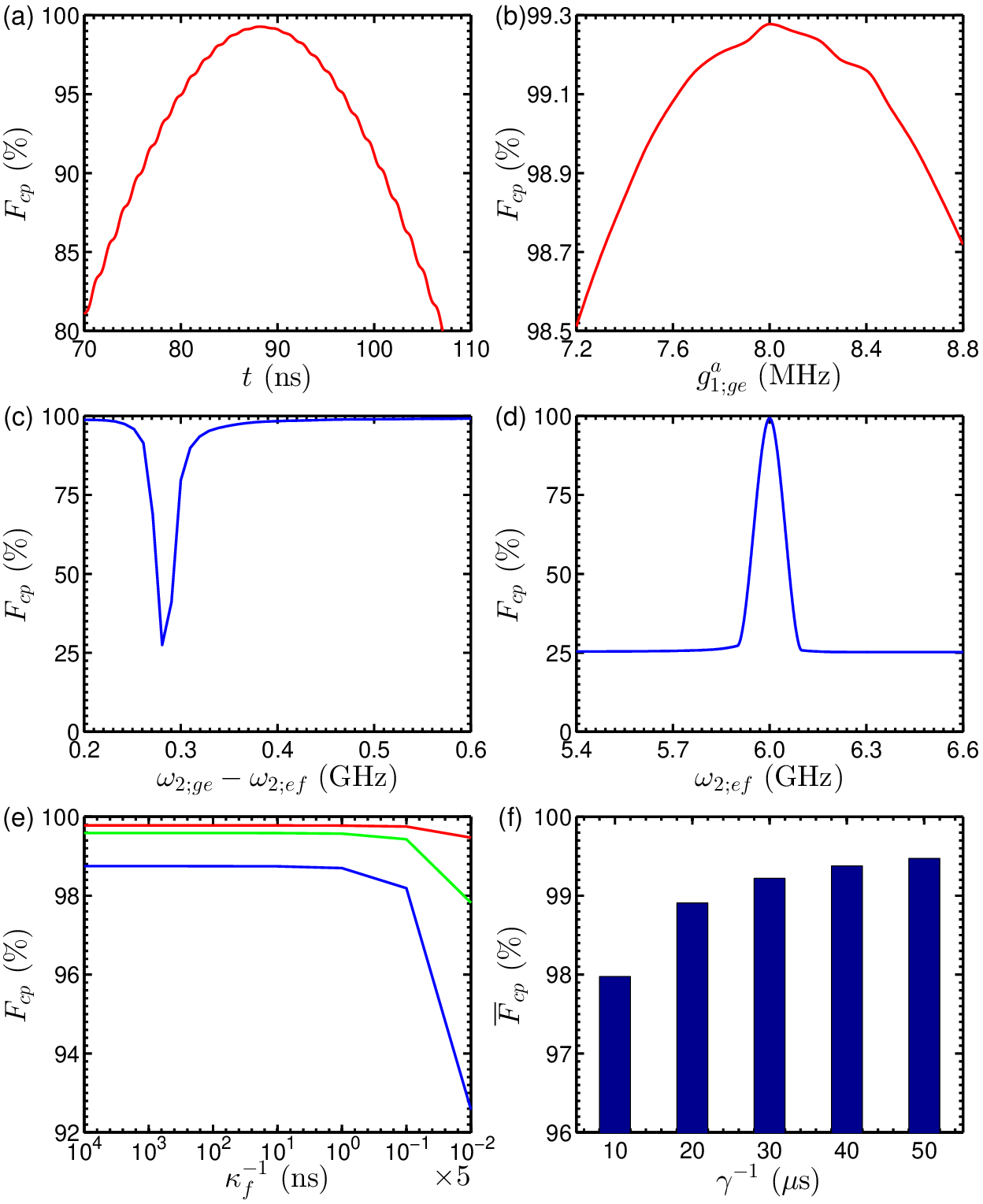}
\caption{(a) The fidelity of the c-phase gate on $q_1$ and $q_2$
varies with the operation time $t$. (b)-(f) The relations between
the fidelity of the gate and $g_{1;ge}^{a}$,
$\omega_{2;ge}-\omega_{2;ef}$, $\omega_{2;ef}$, $\kappa_{f}^{-1}$,
and $\gamma^{-1}$, respectively.} \label{fig7}\end{center}
\end{figure}

To show the feasibility of our c-phase gate on remote qutrits $q_1$
and $q_2$ with decoherence time and the decay time of the qutrits
and the resonators, we numerically simulate the fidelity of
$|\Psi_{max}^{cp}\rangle$ after our c-phase gate operations on the
whole system (the initial state of the system is
$|\Psi_{max}\rangle$) by using the definition
\begin{eqnarray}        
F_{cp} = \langle \Psi_{max}^{cp}|\rho(t)|\Psi_{max}^{cp}\rangle,
\label{Fcp}
\end{eqnarray}
in which the effects from the unresonant parts
\begin{eqnarray}          
H_{1} = g_{a,1}^{e,f}(a\sigma
_{1;e,f}^{+}e^{i\delta_{a,1}^{e,f}t}+a^{+}\sigma
_{1;e,f}^{-}e^{-i\delta_{a,1}^{e,f}t})  \label{cp1}
\end{eqnarray}
and
\begin{eqnarray}          
H_{2} = g_{b,2}^{g,e}(b\sigma
_{2;g,e}^{+}e^{i\delta_{b,2}^{g,e}t}+b^{+}\sigma
_{2;g,e}^{-}e^{-i\delta_{b,2}^{g,e}t})  \label{cp2}
\end{eqnarray}
are considered. Parameters chosen here are
$\omega_{1;ge}/(2\pi)=\omega_{2;es}/(2\pi)=\omega_{a}/(2\pi)=\omega_{b}/(2\pi)=\omega_{f}/(2\pi)=
6$ GHz, $\sqrt{\frac{2}{3}}g_{2;ge}^{b}/(2\pi)=g_{1;ge}^{a}/(2\pi)=
8$ MHz. $\gamma_{l;ge}^{-1}=\kappa_{a,b,f}^{-1}= 50$ $\mu$s. As
shown in Fig. \ref{fig7} (a), the fidelity of the state
$|\Psi_{max}^{cp}\rangle$ can reach $99.28\%$ within $88.1$ ns.

In the realistic experiment, parameters of the system can not match
the ones accurately chosen above. We give the influences on the
fidelity of the state $|\Psi_{max}^{cp}\rangle$ from the coupling
strength, the anharmonicity, the decoherence time, and the frequency
of the qutrits and the decay time of resonators as shown in Fig.
\ref{fig7} (b)-(f). In each figure in Fig. \ref{fig7}, parameters
are kept unchanged except for the one signed in the axis of
abscissas. The influences from $g_{1;ge}^{a}$ is shown in Fig.
\ref{fig7} (b) and the small change of $g_{1;ge}^{a}$ influences the
fidelity little. Fig. \ref{fig7} (c) indicates that the
anharmonicity of $q_2$ should be chosen to let the transition
frequency $\omega_{2;ge}$ detune with $\omega_f +
\sqrt{2}g_{f}^{a(b)}$ largely, which is required when we reduce the
Hamiltonian from $H_{2q}$ to $H_{e\!f\!f}''$. Because of the four-step coupling between the two qutrits, accurate resonance between the two remote qutrits is required as shown in Fig.
\ref{fig7} (d). In the large-scale integration of our system, the
interaction between qutrits can be turned off conveniently by tuning
the frequencies of the qutrits. In Fig. \ref{fig7} (e), we give the
influences on the fidelity of the state from the decay time of
$r_f$. It can be seen that when $\kappa^{-1}_{f}>10$ ns,
$\kappa_{f}^{-1}$ influences the fidelity of the state little. To
show the possible influence from the decay rates of the resonators
and the decoherence time of the qutrits, we calculate the average
gate fidelity of the c-phase gate with different
$\Gamma^{-1}=\gamma_{l;ge}^{-1}=\kappa_{a,b,f}^{-1}$, shown in Fig.
\ref{fig7} (f) by using the average-gate-fidelity definition
\begin{eqnarray}          
F = (\frac{1}{2\pi})^2 \int_0^{2\pi} \int_0^{2\pi} \langle
\Psi_{f}^{cp}|\rho(t)|\Psi_{f}^{cp}\rangle d\theta_1 d\theta_2.
\label{fidelity}
\end{eqnarray}
Here, $\rho(t)$ is the realistic density operator after our c-phase
gate operation on the initial state $\Psi_{0}^{cp}$ with the
Hamiltonian $H$. It is worth noticing that the decay time $20$
$\mu$s corresponds to the typically quality factor $Q \sim
7\times10^5$ of a 1D superconducting resonator \cite{M} and the
coherence time $20$ $\mu$s of a transmon qutrit can also be readily
realized in experiment \cite{Chow}. Although the coupling strength
$g_{f}^{a(b)}/(2\pi)$ taken here is $200$ MHz (predicted
theoretically in \cite{Peropadre}) which satisfies $\Delta=25$ and
it has not been realized in experiments, if we take $\Delta=5$
($g_{f}^{a(b)}/(2\pi)=40$ MHz) and the more of the actual situation
of the life time of SRs and qutrits with $\Gamma^{-1}=20$ $\mu$s,
the fidelity of our c-phase gate can also reach a high value of
$98\%$ (compared with the fidelities between $\Delta=25$ and
$\Delta=5$ as shown in Fig. \ref{fig5}) which should be enhanced
further by taking corresponding optimal parameters. Corresponding to
the operation time of the c-phase gate construction, the length of
the superconducting TL can, in principle, reach the scale of several
meters.

\section{Discussion and summary}

On one hand, in order to use the dark photons in the TL to achieve
the c-phase gate on qutrits $q_1$ and $q_2$, one should take
$g_{1(2);ge}^{a(b)} \ll g_{f}^{a(b)}$. On the other hand, the small
coupling strength of $g_{1(2);ge}^{a(b)}$ used here compared with anharmonicites of transmon qutrits (hundreds of megahertz) does not require the large
anharmonicities of the qutrits. Moreover, to achieve our c-phase
scheme, the $\Xi$-type energy level of the qutrits is required.
Besides the transmon qutrit, the superconducting charge qutrit
\cite{Nakamura,Shnirman} or phase qutrit \cite{Martinis} can also
been applied to our scheme. By using the transmon qutrit or the
phase qutrit with $\omega_{ge}/(2\pi)>\omega_{ef}/(2\pi)$, one
should take the proper anharmonicity of $q_2$ to let the transition
frequency $\omega_{2;ge}$ detune with $\omega_f +
\sqrt{2}g_{f}^{a(b)}$ largely. By using the charge qutrit with
$\omega_{ge}/(2\pi)<\omega_{ef}/(2\pi)$, one should take
$\omega_{2;ge}$ detune with $\omega_f - \sqrt{2}g_{f}^{a(b)}$
largely. That is, when the frequency $\omega_{2;ge} \sim \omega_f +
\sqrt{2}g_{f}^{a(b)}$, the effective Hamiltonians $H_{e\!f\!f}$ and
$H_{e\!f\!f}'$ can not be obtained as the mode $ C_{\pm}$ can not be
suppressed effectively.

In summary, we have proposed a one-step scheme to achieve the
c-phase gate on two remote transmon qutrits coupled to different
resonators connected by a superconducting TL. The scheme works in
the all-resonance regime with just one step, which leads to a fast
operation and can be demonstrated in experiment easily. Moreover,
our scheme is robust against the TL loss by using the dark photon.
That is, the superconducting TL needs not to be populated, which is
convenient to be extended to a large-scale integration condition by
the complex designation of a long-length TL to link lots of remote
circuit QEDs.

\section*{Acknowledgments}

F.G. Deng was supported by the National Natural Science Foundation
of China under Grants No. 11474026, No. 11647042, and No. 11674033,
the Fundamental Research Funds for the Central Universities under
Grant No. 2015KJJCA01. M. Hua was supported by the National Natural
Science Foundation of China under Grants No. 11647042 and No.
11704281.



\begin{thebibliography}{61}



\bibitem{Nielsen} M. A. Nilsen and I. L. Chuang,
\emph{Quantum Computation and Quantum Information} (Cambridge University, Cambridge, 2000).




\bibitem{longguilu} G. L. Long, Commun. Theor. Phys. \textbf{45}, 825 (2006).




\bibitem{shor} P. W. Shor,  SIAM J. Sci. Statist.
Comput.   \textbf{26}, 1484  (1997).



\bibitem{Grover} L. K. Grover, Phys. Rev. Lett.   \textbf{79}, 325  (1997).



\bibitem{LongGrover} G. L. Long,
 Phys. Rev. A  \textbf{64}, 022307  (2001).




\bibitem{Knill} E. Knill, R. Laflamme,  and G. J. Milburn,  Nature
\textbf{409}, 46  (2001).



\bibitem{photon} J. L. O'Brien, G. J. Pryde, A. G. White, T. C. Ralph, and D. Branning,
Nature   \textbf{426},  264 (2003).


\bibitem{hyperCNOT2} B. C. Ren and F. G. Deng,    Sci. Rep.   \textbf{4},
4623 (2014).


\bibitem{hyperCNOT3} B. C. Ren, G. Y. Wang, and F. G. Deng,   Phys. Rev. A   \textbf{91}, 032328 (2015).

\bibitem{hyperCNOT4} B. C. Ren, and F. G. Deng,  Opt. Express \textbf{25}, 10863 (2017).

\bibitem{hyperCNOT5} T. Li,  and G. L. Long, Phys. Rev. A \textbf{94}, 022343 (2016).



\bibitem{photon2} H. R. Wei and F. G. Deng,   Opt. Express   \textbf{21},  17671  (2013).



\bibitem{NMR} J. A. Jones, M. Mosca and  R. H. Hansen,    Nature
 \textbf{393},  344 (1998).


\bibitem{Long1} G. L. Long and L. Xiao,
J. Chem. Phys.    \textbf{119},  8473 (2003).


\bibitem{NHQCLong} G. R. Feng, G. F. Xu,
  and  G. L. Long,   Phys. Rev. Lett.  \textbf{110},   190501 (2013).



\bibitem{Togan} E. Togan, Y. Chu, A. S. Trifonov, L. Jiang, J. Maze, L. Childress,
M. V. G. Dutt, A. S. S\o rensen, P. R. Hemmer, A. S. Zibrov, and  M.
D. Lukin,   Nature    \textbf{466},  730 (2010).


\bibitem{NVgate} W. L. Yang, Z. Q. Yin, Z. Y. Xu, M. Feng, and  J. F. Du,
 Appl. Phys. Lett.  \textbf{96}, 241113  (2010).


\bibitem{weiNVgate} H. R. Wei and F. G. Deng,   Phys. Rev. A   \textbf{88},  042323 (2013).

\bibitem{songxk1} X. K. Song, H. Zhang, Q. Ai, J. Qiu, and F. G.
Deng, New J. Phys. \textbf{18}, 023001 (2016).

\bibitem{songxk2} X. K. Song, Q. Ai, J. Qiu, and F. G. Deng, Phys. Rev. A \textbf{93}, 052324
(2016).


\bibitem{cai} K. Cai, R. X. Wang, Z. Q. Yin, and G. L. Long, Sci. China-Phys. Mech. Astron. \textbf{60}, 070311 (2017).




\bibitem{book} M. O. Scully and M. S. Zubairy,
\emph{Quantum Optics} (Cambridge University, Cambridge, 1997).



\bibitem{Blais} A. Blais, R. S. Huang, A. Wallraff, S. M. Girvin,
and R. J. Schoelkopf,    Phys. Rev. A   \textbf{69},  062320 (2004).


\bibitem{Wallraff} A. Wallraff, D. I. Schuster, A. Blais, L. Frunzio,
R. S. Huang, J. Majer, S. Kumar, S. M. Girvin, and  R. J.
Schoelkopf,  Nature   \textbf{431}, 162 (2004).



\bibitem{Barends} R. Barends, J. Kelly, A. Megrant, A. Veitia, D. Sank, E. Jeffrey,
T. C. White, J. Mutus, A. G. Fowler, B. Campbell, Y. Chen, Z. Chen,
B. Chiaro, A. Dunsworth, C. Neill, P. O'Malley, P. Roushan, A.
Vainsencher, J. Wenner, A. N. Korotkov, A. N. Cleland,  J. M.
Martinis,  Nature    \textbf{508},  500  (2014).



\bibitem{zhengyuan} Z. Y. Xue, Z. Q. Yin, Y. Chen, Z. D. Wang, and S. L. Zhu, Sci. China-Phys. Mech. Astron. \textbf{59} 660301 (2016).




\bibitem{DiCarlo} L. DiCarlo, J. M. Chow, J. M. Gambetta, L. S. Bishop,
B. R. Johnson, D. I. Schuster, J. Majer, A. Blais, L. Frunzio, S. M.
Girvin,  R. J. Schoelkopf,    Nature   \textbf{460}, 240  (2009).



\bibitem{Haack} G. Haack, F. Helmer, M. Mariantoni, F. Marquardt, and E. Solano,
Phys. Rev. B   \textbf{82}, 024514  (2010).



\bibitem{Strauch} F. W. Strauch,
 Phys. Rev. A   \textbf{84}, 052313  (2011).



\bibitem{Hua1} M. Hua,  M. J. Tao, and  F. G. Deng,
Phys. Rev. A    \textbf{90},  012328 (2014).


\bibitem{Hua2} M. Hua, M. J. Tao, and  F. G. Deng,    Sci. Rep.  \textbf{5}, 9274  (2015).




\bibitem{McKay} D. C. McKay, R. Naik, P. Reinhold, L. S. Bishop, and  D. I. Schuster,
Phys. Rev. Lett.    \textbf{114},  080501  (2015).




\bibitem{HPaik} H. Paik, A. Mezzacapo, M. Sandberg, D. T. McClure,
B. Abdo, A. D. C\'orcoles, O. Dial, D. F. Bogorin, B. L. T. Plourde,
M. Steffen, A. W. Cross, J. M. Gambetta, and J. M. Chow,    Phys.
Rev. Lett. \textbf{117},  250502 (2016).




\bibitem{Steffen} M. Steffen, M. Ansmann, R. C. Bialczak,
N. Katz, E. Lucero, R. McDermott, M. Neeley, E. M. Weig, A. N.
Cleland, and J. M. Martinis,
 Science  \textbf{313}, 1423  (2006).



\bibitem{Cao} Y. Cao, W. Y. Huo, Q. Ai, and G. L. Long,
 Phys. Rev. A  \textbf{84}, 053846  (2011).



\bibitem{Leghtas} Z. Leghtas, U. Vool, S. Shankar, M. Hatridge,
S. M. Girvin, M. H. Devoret, and  M. Mirrahimi,  Phys. Rev. A
\textbf{88}, 023849 (2013).



\bibitem{Strauch1} F. W. Strauch,  Phys. Rev. Lett.    \textbf{109}, 210501  (2012).



\bibitem{Strauch2} F.W. Strauch, D. Onyango, K. Jacobs, R. W. Simmonds,
Phys. Rev. A   \textbf{85},  022335 (2012).



\bibitem{Aron} C. Aron, M. Kulkarni, and H. E. T\"ureci,
 Phys. Rev. A   \textbf{90}, 062305  (2014).




\bibitem{Felicetti} S. Felicetti, M. Sanz, L. Lamata, G. Romero,
G. Johansson, P. Delsing, and  E. Solano,    Phys. Rev. Lett.
\textbf{113},  093602(2014).



\bibitem{Narla} A. Narla, S. Shankar, M. Hatridge, Z. Leghtas,
K. M. Sliwa, E. Zalys-Geller, S. O. Mundhada, W. Pfaff, L. Frunzio,
R. J. Schoelkopf,   M. H. Devoret,   Phys. Rev. X   \textbf{6},
 031036 (2016).




\bibitem{Koshino} K. Koshino, K. Inomata, Z. R. Lin, Y. Tokunaga,
T. Yamamoto, and Y. Nakamura,   arXiv: 1610.02104 (2016).




\bibitem{AWallraff} A. Wallraff, D. I. Schuster,
A. Blais, L. Frunzio, J. Majer, M. H. Devoret, S. M. Girvin, and  R.
J. Schoelkopf,   Phys. Rev. Lett.   \textbf{95},  060501 (2005).



\bibitem{Johnson} B. R. Johnson, M. D. Reed, A. A. Houck,
D. I. Schuster, L. S. Bishop, E. Ginossar, J. M. Gambetta, L.
DiCarlo, L. Frunzio, S. M. Girvin, and  R. J. Schoelkopf,   Nat.
Phys. \textbf{6}, 663 (2010).



\bibitem{Feng} W. Feng, P. Y. Wang, X. M. Ding, L. T. Xu,  and X. Q. Li,
Phys. Rev. A  \textbf{83}, 042313  (2011).



\bibitem{hantianyi} T. Y. Han, M. B. Chen, G. Cao, H. O. Li, M. Xiao, and G. P. Guo, Sci. China-Phys. Mech. Astron. \textbf{60}, 057301 (2017).




\bibitem{Majer} J.  Majer, J. M. Chow, J. M. Gambetta, J. Koch, B. R. Johnson, J. A. Schreier,
 L. Frunzio,  D. I. Schuster,
A. A. Houck, A. Wallraff, A. Blais, M. H. Devoret, S. M. Girvin, and
R. J. Schoelkopf,  Nature   \textbf{449},  443 (2007).



\bibitem{Hua3} M. Hua, M. J. Tao, and F. G. Deng,
 Sci. Rep.   \textbf{6}, 22037  (2016).



\bibitem{Wu} C. W. Wu, M. Gao, H. Y. Li, Z. J. Deng, H. Y. Dai, P. X. Chen, and  C. Z. Li,
Phys. Rev. A  \textbf{85}, 042301  (2012).



\bibitem{Yang} C. P. Yang, Q. P. Su, and S. Y. Han,
 Phys. Rev. A    \textbf{86},
022329 (2012).




\bibitem{YangCP} C. P. Yang, Q. P. Su, S. B. Zheng,  and F. Nori,
 Phys. Rev. A    \textbf{93}, 042307  (2016).



\bibitem{Pellizzari} T. Pellizzari,
 Phys. Rev. Lett.   \textbf{79}, 5242  (1997).



\bibitem{Loo} A. F. van Loo, A. Fedorov, K. Lalumi\'ere, B. C. Sanders,
A. Blais, and  A. Wallraff,   Science  \textbf{342}, 1494  (2013).



\bibitem{YY} Y. Yin, Y. Chen, D. Sank, P. J. J. O'Malley,
T. C. White, R. Barends, J. Kelly, E. Lucero, M. Mariantoni, A.
Megrant, C. Neill, A. Vainsencher, J. Wenner, A. N. Korotkov, A. N.
Cleland, J. M. Martinis,   Phys. Rev. Lett.    \textbf{110}, 107001
(2013).



\bibitem{SJ} S. J. Srinivasan, N. M. Sundaresan, D. Sadri, Y. Liu,
J. M. Gambetta, T. Yu, S. M. Girvin, and  A. A. Houck,  Phys. Rev. A
\textbf{89}, 033857 (2014).




\bibitem{MP} M. Pechal, L. Huthmacher, C. Eichler, S. Zeytino\v{g}lu,
A. A. Abdumalikov, J. S. Berger, A. Wallraff, and S. Filipp,
  Phys. Rev. X    \textbf{4},
041010 (2014).




\bibitem{Roch} N. Roch, M. E. Schwartz, F. Motzoi, C. Macklin, R. Vijay, A. W. Eddins,
A. N. Korotkov, K. B. Whaley, M. Sarovar, and I. Siddiqi,  Phys.
Rev. Lett.  \textbf{112}, 170501  (2014).



\bibitem{Mandt} S. Mandt, D. Sadri, A. A. Houck, and H. E. T\"ureci,
 New J. Phys.  \textbf{17}, 053018  (2015).



\bibitem{sheng} Y. B. Sheng, and L. Zhou, Sci. Bull. \textbf{62}, 1025 (2017).




\bibitem{qiang} X. G. Qiang, X. Q. Zhou, K. Aungskunsiri, H. Cable, and J. LO'Brien, Quantum Sci. Technol. \textbf{2},  045002 (2017).




\bibitem{zhoulan1} L. Zhou and Y. B. Sheng, Ann. Phys. \textbf{385}, 10 (2017).




\bibitem{zhoulan2} L. Zhou and Y. B. Sheng, Phys. Rev. A \textbf{92}, 042314 (2015).




\bibitem{zhoulan3} L. Zhou and Y. B. Sheng, Phys. Rev. A \textbf{90}, 024301 (2014).





\bibitem{yanga} G. A. Yan, H. X. Qiao, H. Lu, A. X. Chen, Sci. China-Phys. Mecha. Astron. \textbf{60}, 090311 (2017).




\bibitem{Cirac} J. I. Ciracf, A. K. Ekert, S. F. Huelga, and  C. Macchiavello,
Phys. Rev. A  \textbf{59}, 4249  (1999).



\bibitem{Clark} S. Clark, A. Peng, M. Gu, and  S. Parkins,
Phys. Rev. Lett.    \textbf{91},  177901 (2003).



\bibitem{Browne} D. E. Browne, M. B. Plenio,  and  S. F. Huelga,
 Phys. Rev. Lett.     \textbf{91},   067901 (2003).



\bibitem{Duan1} L. M. Duan and H. J. Kimble,  Phys.
Rev. Lett.   \textbf{90},  253601 (2003).



\bibitem{Mancini} S. Mancini and S. Bose,   Phys. Rev. A   \textbf{70},
022307 (2005).




\bibitem{dengfuguo} F. G. Deng, B. C. Ren, and X. H. Li, Sci. Bull. \textbf{62} 46 (2017).





\bibitem{yanhui} H. Yan, and J. F. Chen, Sci. China-Phys. Mech. Astron. \textbf{58}, 074201 (2015).





\bibitem{Cirac1} J. I. Cirac, P. Zoller, H. J. Kimble, and H. Mabuchi,
 Phys. Rev. Lett.   \textbf{78},
3221 (1997).



\bibitem{Xiao} Y. F. Xiao, X. M. Lin, J. Gao, Y. Yang, Z. F. Han, and G. C. Guo,
Phys. Rev. A   \textbf{70}, 042314  (2004).



\bibitem{Lu} X. Y. L\"u, J. Wu, L. L. Zheng, and  Z. M. Zhan,
Phys. Rev. A   \textbf{83}, 042302  (2011).



\bibitem{Yin1} Z. Q. Yin and F. L. Li,
 Phys. Rev. A  \textbf{75},  012324 (2007).



\bibitem{Clader} B. D. Clader,
Phys. Rev. A   \textbf{90},  012324 (2014).



\bibitem{Yin2} Z. Q. Yin, W. L. Yang, L. Y. Sun, and  L. M. Duan,
 Phys. Rev. A   \textbf{91}, 012333 (2015).





\bibitem{Sai} S. Y. Ye, Z. R. Zhong, and S. B. Zheng,  Phys. Rev. A
 \textbf{77},  014303 (2008).





\bibitem{Enk} S. J. van Enk, H. J. Kimble, J. I. Cirac,  and  P. Zoller,
Phys. Rev. A   \textbf{59}, 2659  (1999).



\bibitem{Serafini} A. Serafini, S. Mancini, and S. Bose,   Phys. Rev.
Lett.  \textbf{96},  010503 (2006).



\bibitem{WLYang} W. L. Yang, Y. Hu, Z. Q. Yin, Z. J. Deng,  and  M. Feng,
Phys. Rev. A    \textbf{83},  022302 (2011).




\bibitem{Reed} M. D. Reed, L. DiCarlo, S. E. Nigg, L. Sun,
L. Frunzio, S. M. Girvin,  and  R. J. Schoelkopf,    Nature
\textbf{482}, 382 (2012).



\bibitem{Megrant} A. Megrant, C. Neill, R. Barends, B. Chiaro, Y. Chen,
 L. Feigl, J. Kelly, E. Lucero,
M. Mariantoni, P. J. J. O'Malley, D. Sank, A. Vainsencher, J.
Wenner, T. C. White, Y. Yin, J. Zhao, C. J. Palmstr\o m, J. M.
Martinis,  A. N. Cleland,   Appl. Phys. Lett.  \textbf{100}, 113510
(2012).



\bibitem{Koch} J. Koch, T. M. Yu, J. Gambetta, A. A. Houck, D. I. Schuster, J. Majer,
 A. Blais, M. H. Devoret, S. M. Girvin, and
R. J. Schoelkopf,   Phys. Rev. A   \textbf{76},  042319 (2007).



\bibitem{Schreier} J.A. Schreier, A. A. Houck, J. Koch, D. I. Schuster,
B. R. Johnson, J. M. Chow, J. M. Gambetta, J. Majer, L. Frunzio, M.
H. Devoret, S. M. Girvin,  and  R. J. Schoelkopf,    Phys. Rev. B
\textbf{77},  180502(R) (2008).




\bibitem{Chow} J. M. Chow, J. M. Gambetta, E. Magesan, D. W. Abraham, A. W. Cross, B. R. Johnson,
N. A. Masluk, C. A. Ryan, J. A. Smolin, S. J. Srinivasan, and   M.
Steffen,
 Nat. Commun.   \textbf{5},  4015 (2014).




\bibitem{Chang} J. B. Chang, M. R. Vissers, A. D. C\'orcoles, M. Sandberg,
J. S. Gao, D. W. Abraham, J. M. Chow, J. M. Gambetta, M. B.
Rothwell, G. A. Keefe, M. Steffen, and  D. P. Pappas,    Appl. Phys.
Lett. \textbf{103}, 012602 (2013).



\bibitem{Steffen1} L. Steffen, Y. Salathe, M. Oppliger, P. Kurpiers,
M. Baur, C. Lang, C. Eichler, G. P. Hellmann, A. Fedorov, and  A.
Wallraff,   Nature    \textbf{500}, 319  (2013).



\bibitem{Hoi} I. C. Hoi, C. M. Wilson, G. Johansson, T. Palomaki,
B. Peropadre, and  P. Delsing,   Phys. Rev. Lett.   \textbf{107},
 073601 (2011).



\bibitem{Peropadre} B. Peropadre, D. Zueco, F. Wulschner, F. Deppe, A. Marx,
R. Gross, and  J. J. G. Ripoll,   Phys. Rev. B    \textbf{87},
 134504 (2013).



\bibitem{hu} Y. Hu  and L. Tian,   Phys.
Rev. Lett.   \textbf{106}, 257002  (2011).




\bibitem{Strand} J. D. Strand, M. Ware, F. Beaudoin,
T. A. Ohki, B. R. Johnson, A. Blais, and  B. L. T. Plourde,  Phys.
Rev. B  \textbf{87},  220505(R) (2013).



\bibitem{M} M. G\"oppl, A. Fragner, M. Baur, R. Bianchetti, S. Filipp,
J. M. Fink, P. J. Leek, G. Puebla, L. Steffen, and  A. Wallraff,
  J. Appl. Phys.  \textbf{104}, 113904 (2008).



\bibitem{Nakamura} Y. Nakamura, Y. Pashkin, and J. S. Tsai,
  Nature    \textbf{398}, 786  (1999).



\bibitem{Shnirman} A. Shnirman, G. Sch\"on, and Z. Hermon,
   Phys. Rev.  Lett.   \textbf{79}, 2371  (1997).



\bibitem{Martinis} J. M. Martinis, Quantum Inf. Proc.   \textbf{8},  81 (2009).




%
%
%
%
%
%







\end{thebibliography}
\end{document}